\documentclass[aps,prb,twocolumn,showpacs,superscriptaddress,floatfix]{revtex4}
\usepackage{bm}
\usepackage{color}
\usepackage{amssymb}
\usepackage{graphicx}
\usepackage{amsmath}
\usepackage{dcolumn}
\usepackage{bm}
\usepackage[german,english]{babel}

\newcommand{\beq}{\begin{eqnarray}}
\newcommand{\eeq}{\end{eqnarray}}
\newcommand{\beqa}{\begin{equation}}
\newcommand{\eeqa}{\end{equation}}

\begin{document}

\title{Electronic Properties of Disordered Graphene Antidot Lattices}
\author{Shengjun Yuan}
\email{s.yuan@science.ru.nl}
\affiliation{Radboud University of Nijmegen, Institute for
Molecules and Materials, Heijendaalseweg 135, 6525 AJ Nijmegen,
The Netherlands}
\author{Rafael Rold\'an}
\email{rroldan@icmm.csic.es}
\affiliation{Instituto de Ciencia de Materiales de Madrid, CSIC,
Cantoblanco E28049 Madrid, Spain}
\author{Antti-Pekka Jauho}
\affiliation{Center for Nanostructured Graphene (CNG),   DTU Nanotech, Department of Micro- and Nanotechnology,
Technical University of Denmark, DK-2800 Kongens Lyngby, Denmark}
\author{M. I. Katsnelson}
\affiliation{Radboud University of Nijmegen, Institute for
Molecules and Materials, Heijendaalseweg 135, 6525 AJ Nijmegen,
The Netherlands}
\date{\today}

\begin{abstract}
Regular nanoscale perforations in graphene (graphene antidot lattices, GAL) are known to lead to a gap in the energy spectrum, thereby paving a possible way towards many applications.  This theoretical prediction relies on a perfect placement of identical perforations, a situation not likely to occur in the laboratory.  Here, we present a systematic study of the effects of disorder in GALs.  We consider both geometric and chemical disorder, and evaluate the density-of-states as well as the optical conductivity of disordered GALs.  The theoretical method is based on an efficient algorithm for solving the time-dependent Schr{\"o}dinger equation in a tight-binding representation of the graphene sheet [S. Yuan et al., Phys. Rev. B {\bf 82}, 115448 (2010)], which allows us to consider GALs consisting of 6400 $\times$ 6400 carbon atoms.  The central conclusion for all kinds of disorder is that the gaps found for pristine GALs do survive at a considerable amount of disorder, but disappear for very strong disorder. Geometric disorder is more detrimental to gap formation than chemical disorder. The optical conductivity shows a low-energy tail below the pristine GAL band gap due to disorder-introduced transitions.

\end{abstract}

\maketitle

\section{Introduction}

Pristine graphene has no band gap: the conduction and valence bands touch at the K and K'-points of the hexagonal Brillouin zone.  This property, combined with the linear dispersion of the low-energy excitations leads to the spectacular electronic properties that graphene is so famous for\cite{Neto2009,KatslesonBook}.  Nevertheless, the lack of a gap severely hampers many applications where a gap is needed to control the flow of charges.  This feature is further underscored by the phenomenon of Klein tunneling: graphene carriers impinging on a potential barrier may experience reflectionless tunneling thus making their control even more difficult\cite{KNG06,Beenakker2008}.  It is thus natural that many schemes have been proposed to create a gap in graphene: these suggestions include etching extended graphene flakes into nanoribbons\cite{Son2006,Brey2006}, or by considering bilayer graphene in a transverse electric field\cite{McCann2006,CC07}, or by using an external periodic potential to modify the electronic properties so that a gap is formed.  The external periodic potential may be caused by a number of agents, such as periodic gates\cite{Barbier2008,Goor2012} or strain\cite{Low2011}, or adsorption of adatoms in a regular pattern\cite{Sofo2007,Balog2010}, or, as in this work, by a regular nanoperforation of the pristine graphene sheet; this system will be referred to as graphene antidot lattices (GAL)\cite{PP08}.

The design principle\cite{PP08} behind the GAL was inspired by photonic crystals where pass and stop bands for light can be designed by drilling holes in the dielectric medium.  GALs (and their constituents, single holes in graphene\cite{Palacios2008}) have been studied theoretically with a large number of methods, ranging from a continuum description and tight-binding methods\cite{PP08,Burset2011} to fully microscopic DFT calculations\cite{Furst2009,Furst2009b}.  Both electronic\cite{Vanevic2009,Bliokh2009,Rosales2009,Zheng2009} and thermal\cite{Gunst2011,Yan2012,Chang2012} transport properties, as well as optical properties\cite{PP08b,Garm2009} have been discussed.  Symmetry principles determining the existence or non-existence of the gap have been outlined\cite{Rene2011,Ouyang2011}.  Most important, however, is the recent emergence of experimental techniques by which GALs can be fabricated.  These fabrication methods include, e.g., electron-beam etching\cite{Shen2008,Eroms2009,BS11,Gisbers2012}, etch-masks based on self-assembled block co-polymers\cite{Bieri2009,BD10,KG10b}, nanoimprint technology\cite{LB10}, or nanoparticle deposition\cite{Sinitskii2010,Shimizu2012}.  Most experimental papers have focused on the structural aspects, but also a few transport experiments have been reported\cite{Shen2008,Eroms2009,BD10,Shimizu2012,Gisbers2012}. Indeed, transport gaps have been observed but so far they have been associated to disorder induced localization instead of band-structure effects\cite{BD10,Gisbers2012}.  This highlights the importance of studying disorder in GALs: all fabricated structures contain disorder, and one cannot (yet) control the exact geometry of the edges of the etched holes.  It is thus vital to examine the robustness of the band-gaps against disorder, whether it be structural, geometrical or chemical.  A study of this kind presents a serious computational challenge because the systems fabricated in the lab, where unit cells of the order of tens of nanometers can be achieved, are computationally large involving tens of thousands of carbon atoms in the computational cell.  Fully microscopic DFT-based methods cannot presently address such systems, and certain compromises must be made.

In this paper, we perform a systematic study of the electronic properties of disordered GALs in the framework of a tight-binding model in a perforated honeycomb lattice of carbon atoms. We consider the most relevant kinds of disorder for these systems, namely a random deviation of the periodicity and of the radii of the nanoholes from the perfect array, as well as the effect of resonant scatterers in the sample (like vacancies, adatoms, etc.) and the effect of non-correlated and correlated (Gaussian) on-site potentials. Within this scheme, the density of states (DOS) is obtained from a numerical solution of the time-dependent Schr\"odinger equation (TDSE)\cite{YRK10}, and the optical conductivity is calculated by using the Kubo formula for non-interacting electrons.\cite{YRK10,YRRK11}

The paper is organized as follows. In Sec. \ref{Sec:Method} we present the details of the method. The effect of the different kinds of disorder on the DOS and the optical conductivity of a GAL is discussed in Sec. \ref{Sec:Results}. Finally, our main conclusions are summarized in Sec. \ref{Sec:Conclusions}.

\section{Model and method}\label{Sec:Method}

\begin{figure*}[t]
\begin{center}
\mbox{
\includegraphics[width=0.8\columnwidth]{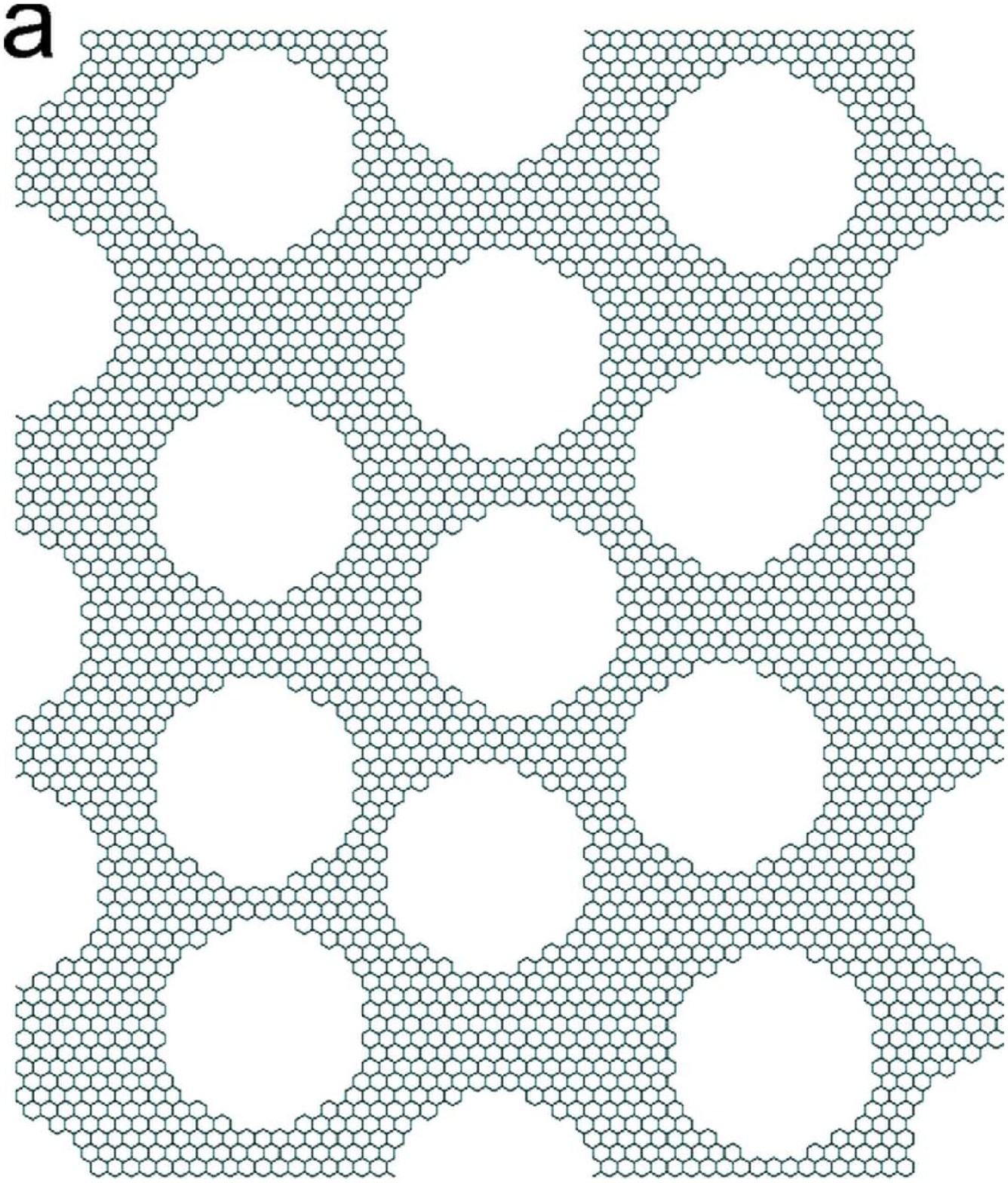}
\includegraphics[width=0.8\columnwidth]{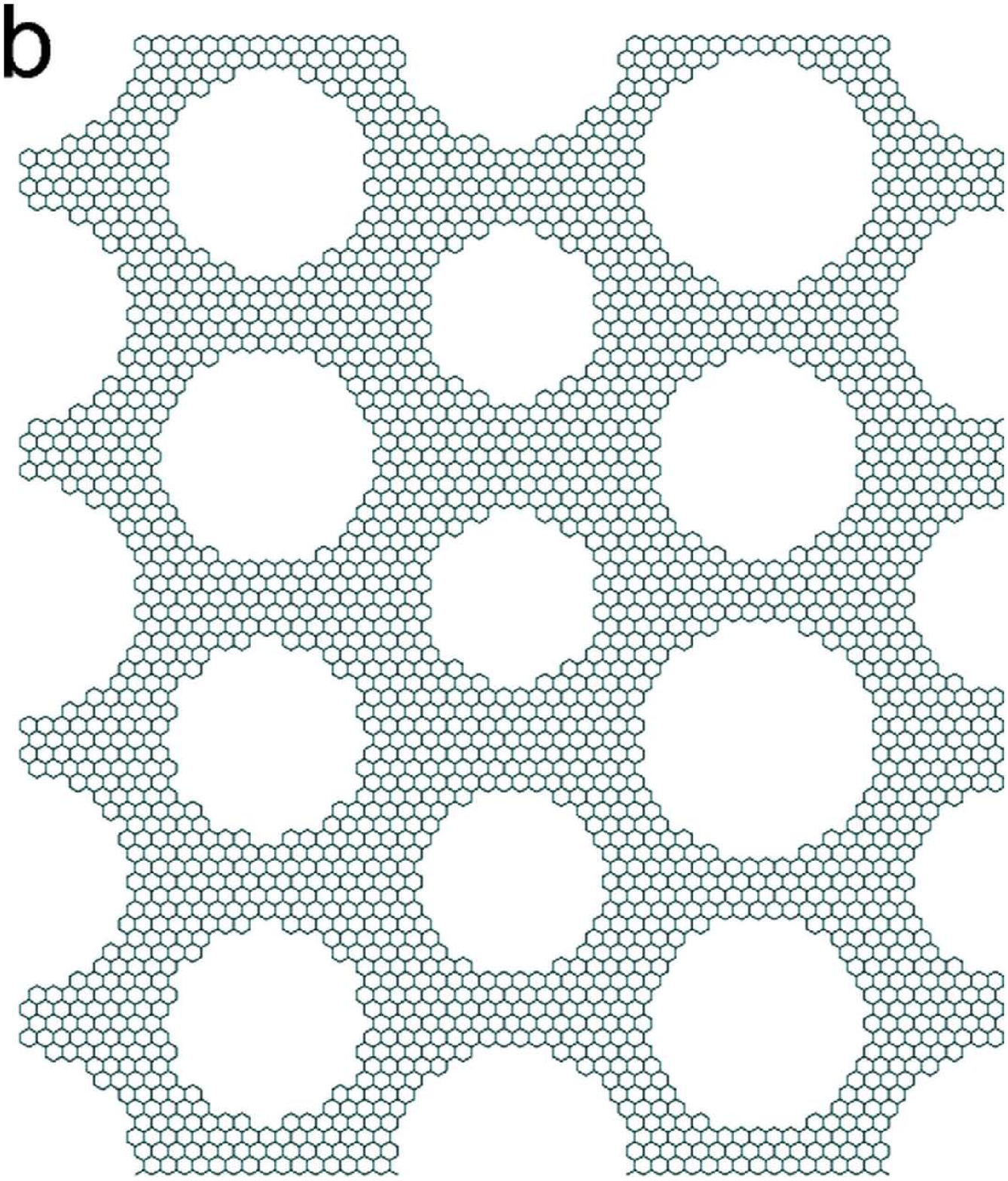}
}
\mbox{
\includegraphics[width=0.8\columnwidth]{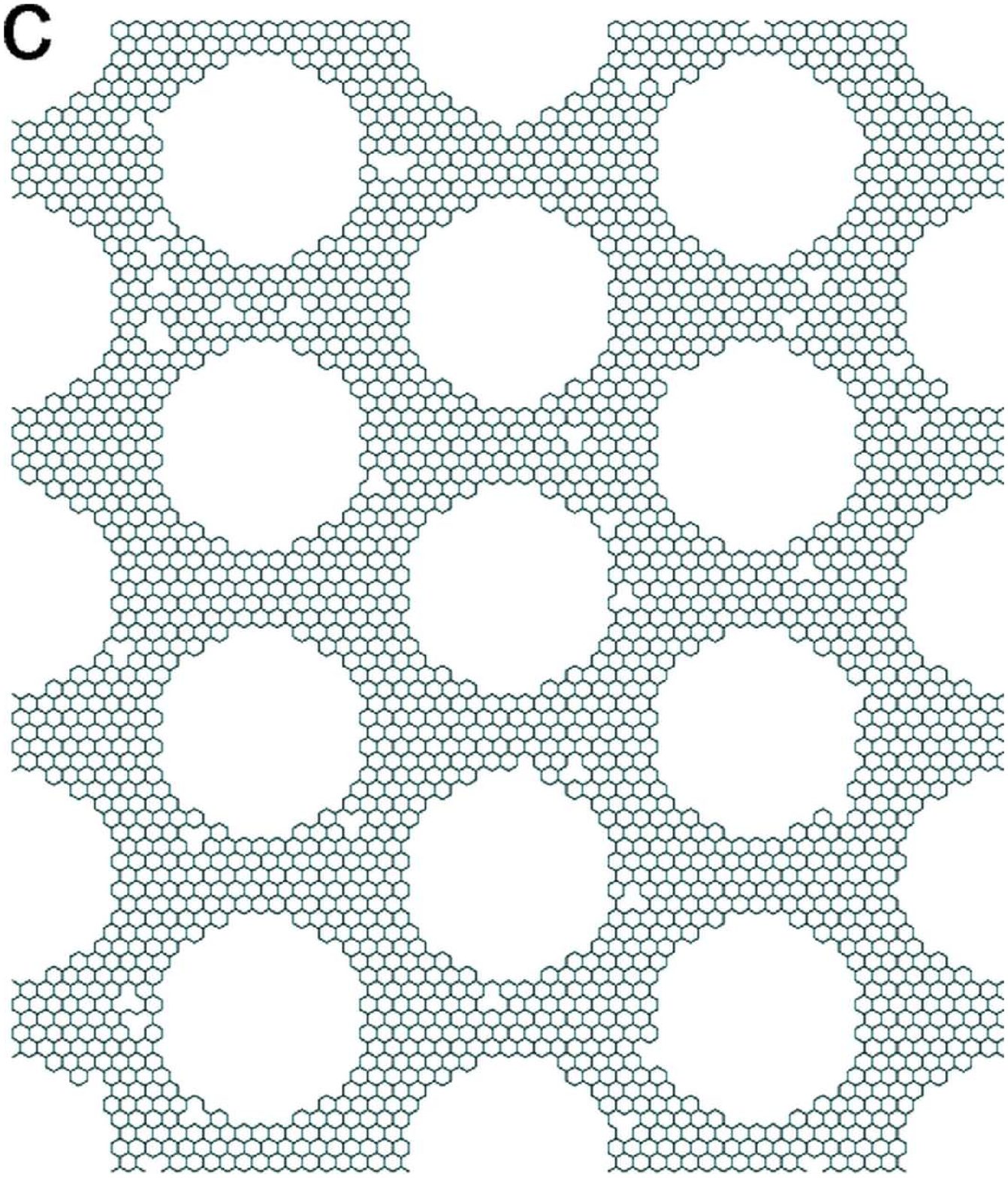}
\includegraphics[width=0.8\columnwidth]{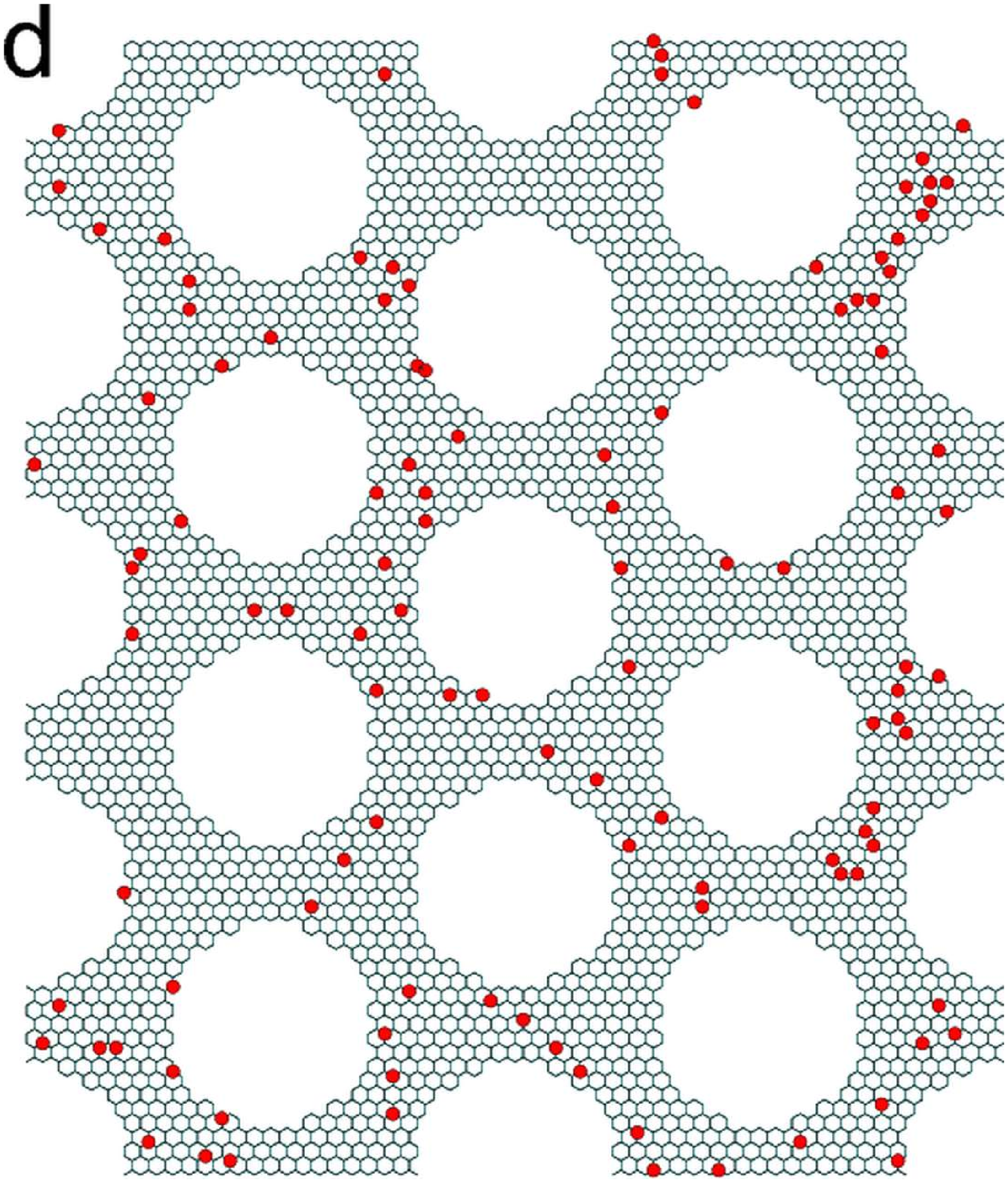}
}
\end{center}
\caption{Sketch of the different kinds of disorder considered. a) The center of the holes is shifted randomly with respect to the original
position in the perfect periodic array $(x,y)$ to a new position in the range $(x\pm l_{C},y\pm l_{C})$ ($l_{C}=2a$). (Notice the different relative distance between the holes) b)
The radius of the holes is randomly shrunk or enlarged within the range $[R-r_{R},R+r_{R}]$ ($r_{R}=a$). (Notice the different relative size of the holes) c) GAL with additional randomly distributed vacancies, signaled by the missing carbon atoms ($n_{x}=1\%$). d) GAL with randomly distributed hydrogen adatoms, signaled by the red dots ($n_{i}=1.75\%$). Notice that two other kinds of disorder are considered in the text, namely non-correlated and correlated long-range (Gaussian) changes in the on-site potentials, which are not sketched in this figure.}
\label{Fig:Sketch}
\end{figure*}

We consider the following real space tight-binding Hamiltonian for a disordered GAL%
\begin{equation}
{\cal H}=-\sum_{\langle i,j\rangle}(t_{ij}a_{i}^{\dagger }b_{j}+\mathrm{h.c})+%
\sum_{i}v_{i}c_{i}^{\dagger }c_{i},+{\cal H}_{imp},  \label{Eq:H0}
\end{equation}%
where $a_{i}^{\dagger }$ ($b_{i}$) creates (annihilates) an electron
on sublattice A (B) of the honeycomb graphene lattice, $t_{ij}$ is the nearest neighbor hopping parameter and $v_{i}$ is the on-site potential. In our model, the GAL is simulated by the creation of an hexagonal array of circular holes of a given radius $R$, and a separation $P=\sqrt{3}L$ between the centers of two consecutive holes, where $L$ is the side length of the hexagonal unit cell.\cite{PP08} We thus label our GALs with the parameters $\{L,R\}$, in units of the graphene lattice constant $a=\sqrt{3}\tilde{a}\approx 2.46$~\AA, where $\tilde{a}\approx 1.42$~\AA~ is the carbon-carbon distance. Another possible notation is $[P,N]$, where $N=\sqrt{3}L-2R$ is the {\it neck width}, defined as the smallest edge-to-edge distance
between two neighbouring holes in the array.\cite{BD10} Deviations of the GAL with respect to perfect periodicity are considered in our calculation in a twofold manner. First, we allow the center of the holes to float with respect to their position in the perfect periodic lattice $(x, y)$ around $(x \pm l_C,y \pm l_C)$ [see Fig. \ref{Fig:Sketch} (a)]. Second, we let the radius of the holes to randomly shrink or widen within the range $[R-r_R, R+r_R]$, as sketched in Fig. \ref{Fig:Sketch} (b). All along this paper, we will express $l_C$ and $r_R$ in units of $a$.

The second term to the right of Eq. (\ref{Eq:H0}) accounts for a change in the on-site potential of the carbon atoms. A long-range potential for correlated impurities can be modelled with
\begin{equation}
v_{i}=\sum_{k=1}^{N_{c}}V_{k}\exp \left( -\frac{\left\vert \mathbf{r}%
_{i}-\mathbf{r}_{k}\right\vert ^{2}}{2d^{2}}\right) ,  \label{Eq:Gaussian}
\end{equation}%
where $N_{c}$ is the number of the impurity centers, which are chosen
randomly distributed on the carbon atoms, $V_{k}$ is uniformly random in the
range $[-V_0 ,V_0 ]$ and $d$ is interpreted as the effective potential
radius. The value of $N_{c}$ is characterized by the ratio $n_{c}=N_{c}/N$, where $N$ is the total
number of carbon atoms of the sample.
A non-correlated short-range random potential can be obtained from the above equation with $d \rightarrow 0$, i.e., $v_i$ is random and uniformly distributed, independently of each site $i$, in the range $[-v_r,+v_r]$. The number of sites with nonzero pontential ($N_{r}$) is characterized as $n_{r}=N_{r}/N$.

We further consider the effect of isolated vacancies in the sample, which can be
regarded as an atom (lattice point) with an on-site energy $%
v_{i}\rightarrow \infty $ or, alternatively, with its hopping amplitudes to other sites
being zero. In the numerical simulation, the simplest way to implement a
vacancy is to remove the atom at the vacancy site [see Fig. \ref{Fig:Sketch} (c)].

If additional resonant impurities are present in the sample as, e. g., hydrogen adatoms, their effect is accounted for through the term ${\cal H}_{imp}$ in Eq. (\ref{Eq:H0}):%
\begin{equation}
{\cal H}_{imp}=\varepsilon _{d}\sum_{i}d_{i}^{\dagger}d_{i}+V\sum_{i}\left(
d_{i}^{\dagger}c_{i}+\mathrm{h.c}\right) ,  \label{Eq:Himp}
\end{equation}%
where $\varepsilon _{d}$ is the on-site potential on the
``hydrogen'' impurity (to be specific, we will use this
terminology although more complicated chemical species can be considered,
such as various organic groups \cite{WK10}) and $V$ is the
hopping between carbon and hydrogen atoms.\cite{Robinson08,WK10,YRK10} The
spin degree of freedom, which contributes through a degeneracy
factor 2, is omitted for simplicity in Eq.~(\ref{Eq:H0}). All along this work, we fix the temperature to
$T=300$K. We use periodic boundary conditions in the calculations for
both the optical conductivity and the density of states, and the
size of the system is $6400\times 6400$ atoms.

Our numerical method is based on an efficient evaluation of the time-evolution operator $e^{-i{\cal H}t}$, based on
the Chebyshev polynomial
representation.\cite{YRK10} (In fact, {\it any} function of $\cal H$ can be evaluated with this method).
We have thus access to the time-dependent state $\left\vert\varphi(t)\right\rangle\equiv
e^{-i{\cal H}t}\left\vert\varphi\right\rangle$, where
$\left\vert
\varphi \right\rangle $ is a random superposition of all the basis
states in the real space, i.e.,\cite{HR00,YRK10}
\begin{equation}
\left\vert \varphi \right\rangle =\sum_{i}a_{i}c_{i}^{\dagger }\left\vert
0\right\rangle ,  \label{Eq:phi0}
\end{equation}
$a_{i}$ are random complex numbers normalized as $\sum_{i}\left\vert
a_{i}\right\vert ^{2}=1$, and $\left\vert0\right\rangle$ is the electron vacuum state.

\begin{figure*}[t]
\begin{center}
\mbox{
\includegraphics[width=0.8\columnwidth]{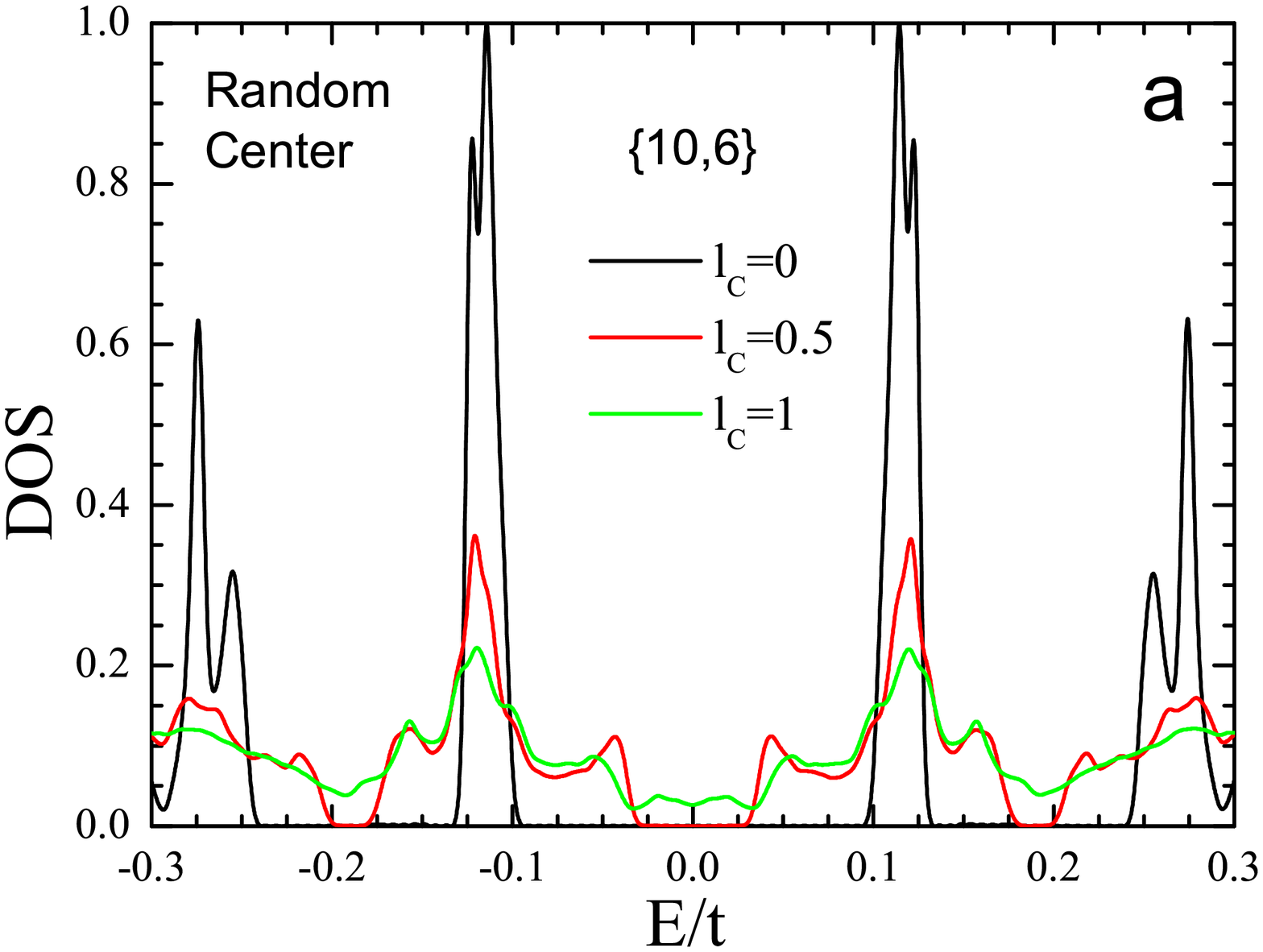}
\includegraphics[width=0.8\columnwidth]{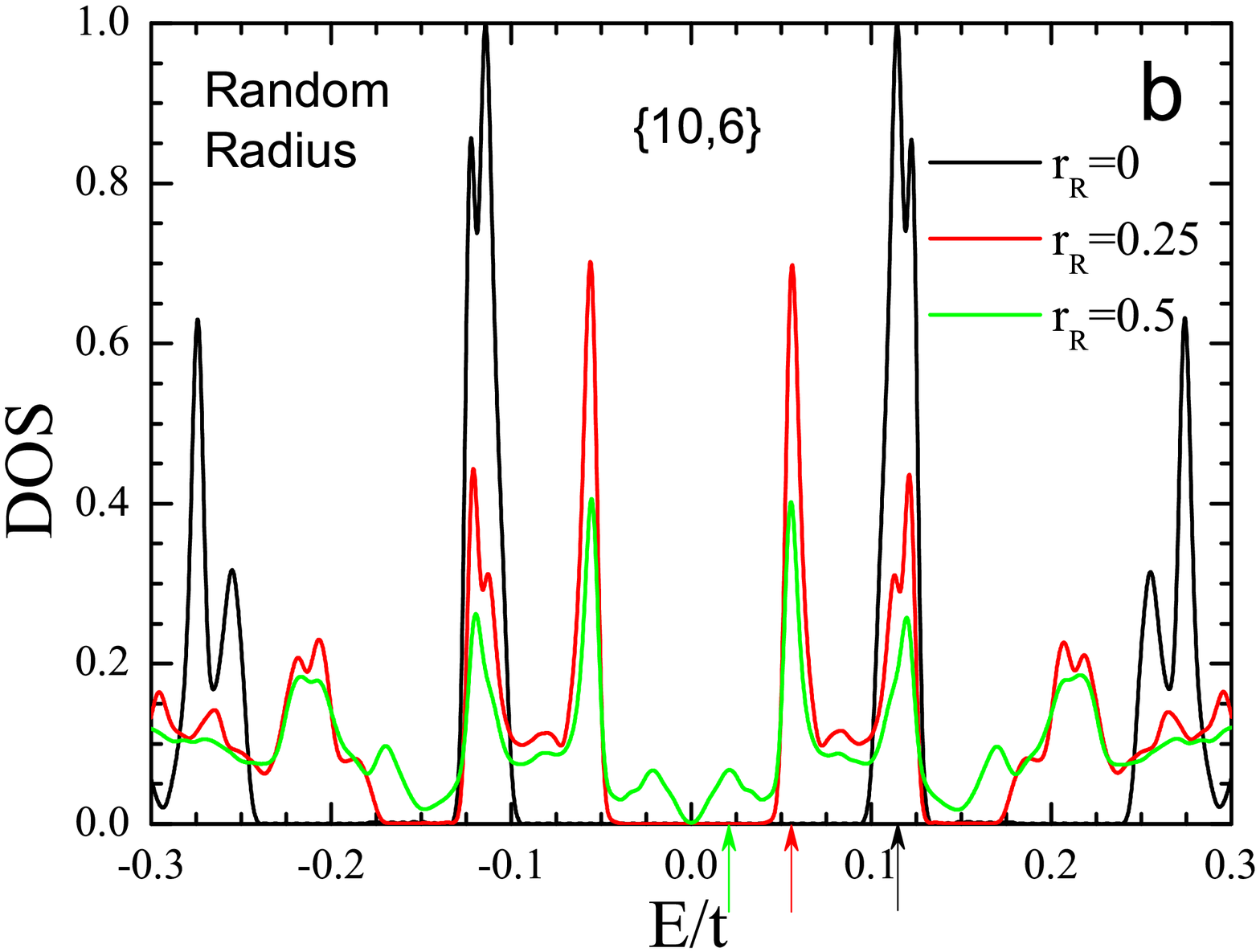}
}
\mbox{
\includegraphics[width=0.8\columnwidth]{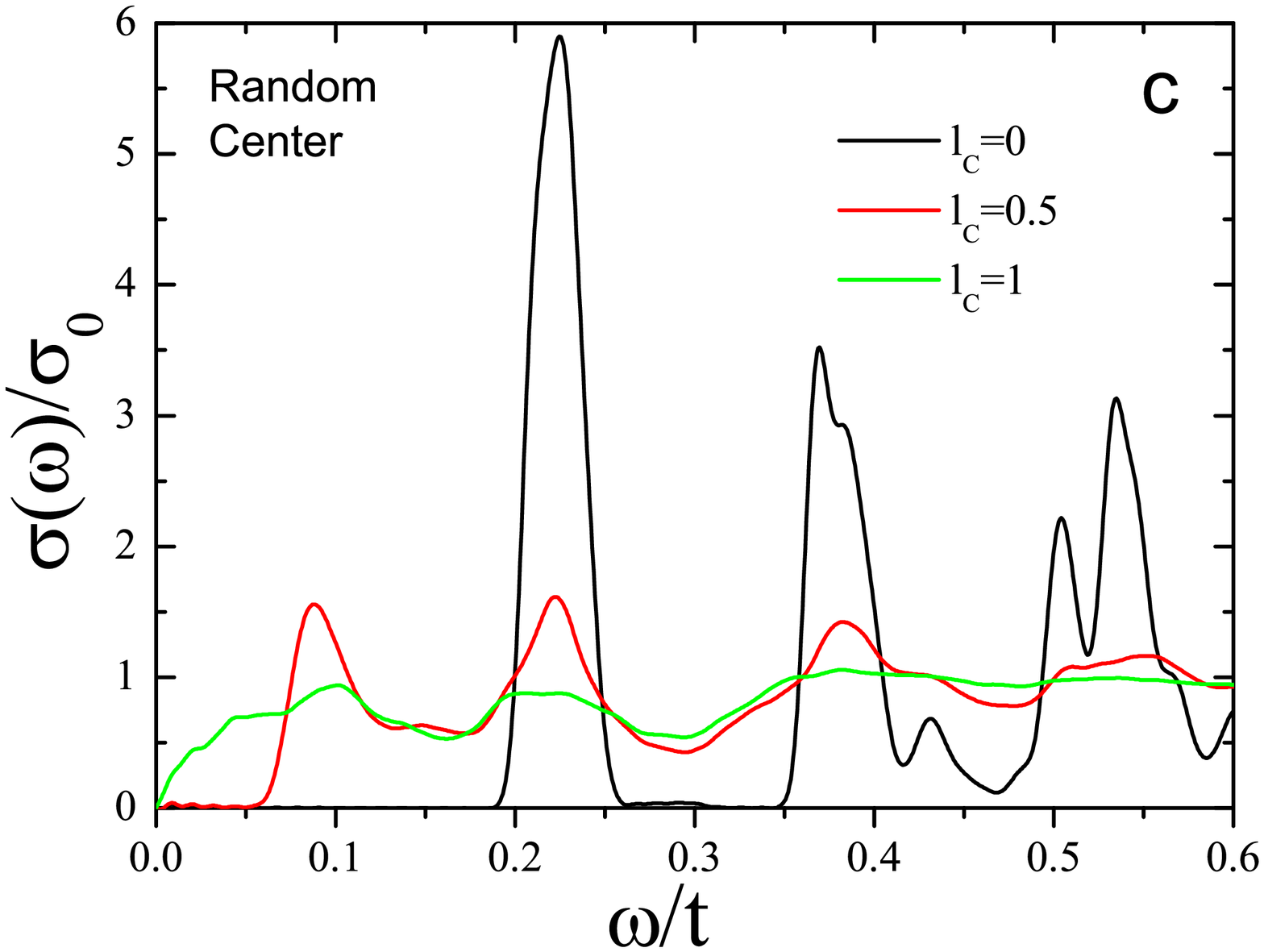}
\includegraphics[width=0.8\columnwidth]{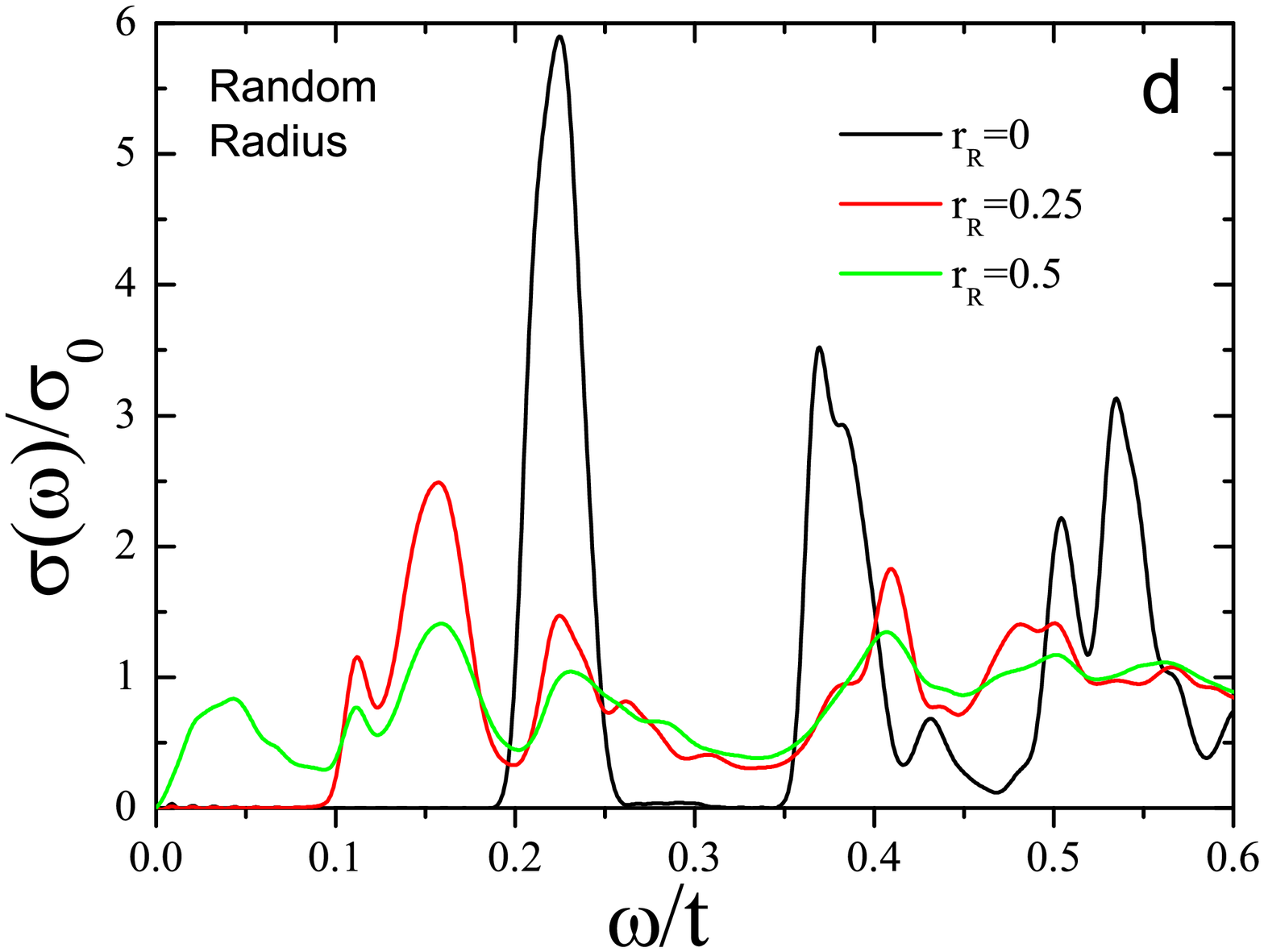}
}
\end{center}
\caption{DOS (top panels) and optical conductivity (bottom panels) for a $\{10,6\}$ GAL with geometrical disorder. In the first column we show (a) DOS and (c) $\sigma(\omega)$ for a disordered GAL in which the center of the holes is shifted randomly with respect to the original
position in the perfect periodic array within the range $(x\pm l_{C},y\pm l_{C})$, as sketched in Fig. \ref{Fig:Sketch} (a). The different colors correspond to different values of $l_C$ (in units of $a$), as denoted in the inset of the figures. In the second column we show (b) DOS and (d) $\sigma(\omega)$ for a GAL where
the radius of the holes is randomly shrunk or enlarged within the range $[R-r_{R},R+r_{R}]$, as sketched in Fig. \ref{Fig:Sketch} (b). Different colors correspond to different values of $r_R$ (in units of $a$).}
\label{Fig:GeomDisorder}
\end{figure*}

 The numerical method has the advantage that an average over different random initial states is not needed. This is because one initial state contains all the eigenstates in the whole spectrum\cite {HR00,YRK10}. Furthermore, it is not necessary to average over different realizations of the disorder, because the system contains millions of carbon atoms, and one specific disordered configuration contains a large number of different local configurations. As shown in Ref.[\onlinecite {YRK10}], the results for different disorder configurations are essentially identical.

Consider first the optical conductivity. We omit in our calculations the $\omega=0$ Drude contribution to the real part of the optical conductivity, so that the regular part can be written as\cite{Ishihara1971,YRK10}
\begin{eqnarray}
\sigma _{\alpha \beta }\left( \omega \right)  &=&\lim_{\varepsilon
\rightarrow 0^{+}}\frac{e^{-\beta \omega }-1}{ \omega \Omega }%
\int_{0}^{\infty }e^{-\varepsilon t}\sin \omega t  \notag  \label{gabw2} \\
&&\times 2\text{Im}\left\langle \varphi |f\left( {\cal H}\right) J_{\alpha }\left(
t\right) \left[ 1-f\left( {\cal H}\right) \right] J_{\beta }|\varphi \right\rangle
dt,  \notag \\
&&
\end{eqnarray}%
where $\beta =1/k_{B}T$ is the inverse
temperature, $\Omega $ is the sample area, $f\left( {\cal H}\right)
=1/\left[ e^{\beta \left( {\cal H}-\mu \right) }+1\right] $ is the
Fermi-Dirac distribution operator, and we use units such that $\hbar=1$. The time-dependent current
operator in the $\alpha $ ($=x$ or $y$) direction is $J_{\alpha }\left( t\right)
=e^{i{\cal H}t}J_{\alpha }e^{-i{\cal H}t}$.   The Fermi-Dirac distribution operator $f\left(
{\cal H}\right) $ is computed with the  Chebyshev polynomial
representation, as mentioned above. 
As the next example, consider the overlap between the time-evolved state  $\left\vert\varphi(t)\right\rangle$ and the initial state $\left\vert\varphi\right\rangle$.  The Fourier transform of this object yields the DOS of the system as\cite{HR00,YRK10}
\begin{equation}
\rho \left( \varepsilon \right) =\frac{1}{2\pi }\int_{-\infty }^{\infty
}e^{i\varepsilon t}\left\langle \varphi \vert
\varphi(t) \right\rangle dt.  \label{Eq:DOS}
\end{equation}

Finally, the quasi-eigenstate $\left\vert \Phi \left( E \right) \right\rangle $, which is a superposition of the degenerate eigenstates with the same eigenenergy $E$, is obtained as the Fourier transform  of $\left\vert\varphi(t)\right\rangle$:\cite{YRK10}%
\begin{equation}
\left\vert \Phi \left( E \right) \right\rangle =\frac{1}{2\pi }%
\int_{-\infty }^{\infty }dte^{iE t}\left\vert \varphi \left(
t\right) \right\rangle .
\end{equation}
The quasi-eigenstate is not exactly an energy eigenstate, unless the
corresponding eigenstate is not degenerate at energy $E$.  However we can
still use the real space distribution of the amplitude to examine the
quasi-localization of the modes\cite{YRK10,YRRK11,Yuan2012}.  Below we display
several examples of all these objects.

\section{Results and discussion}\label{Sec:Results}

\begin{figure*}[t]
\begin{center}
\mbox{
\includegraphics[width=\columnwidth]{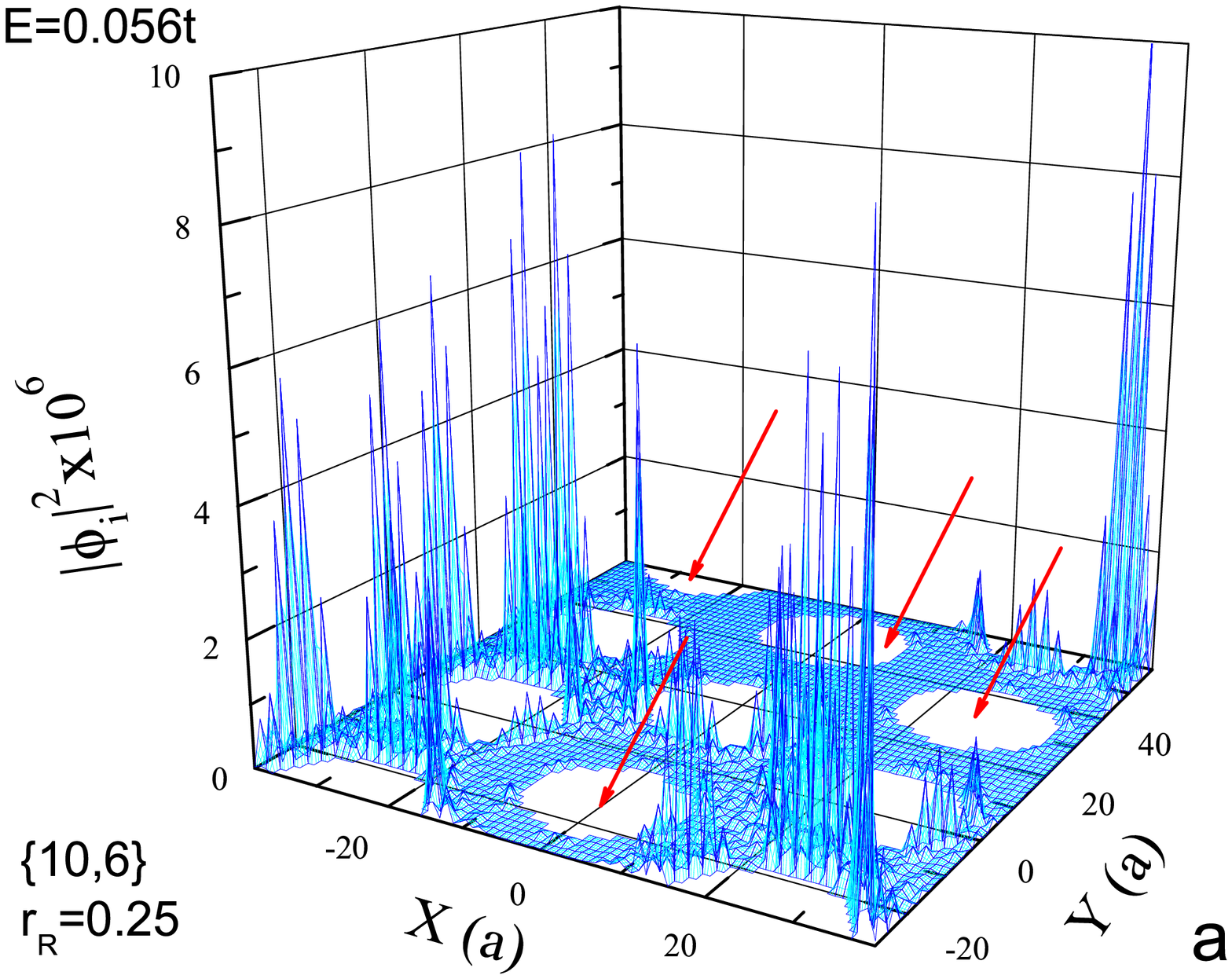}
\includegraphics[width=\columnwidth]{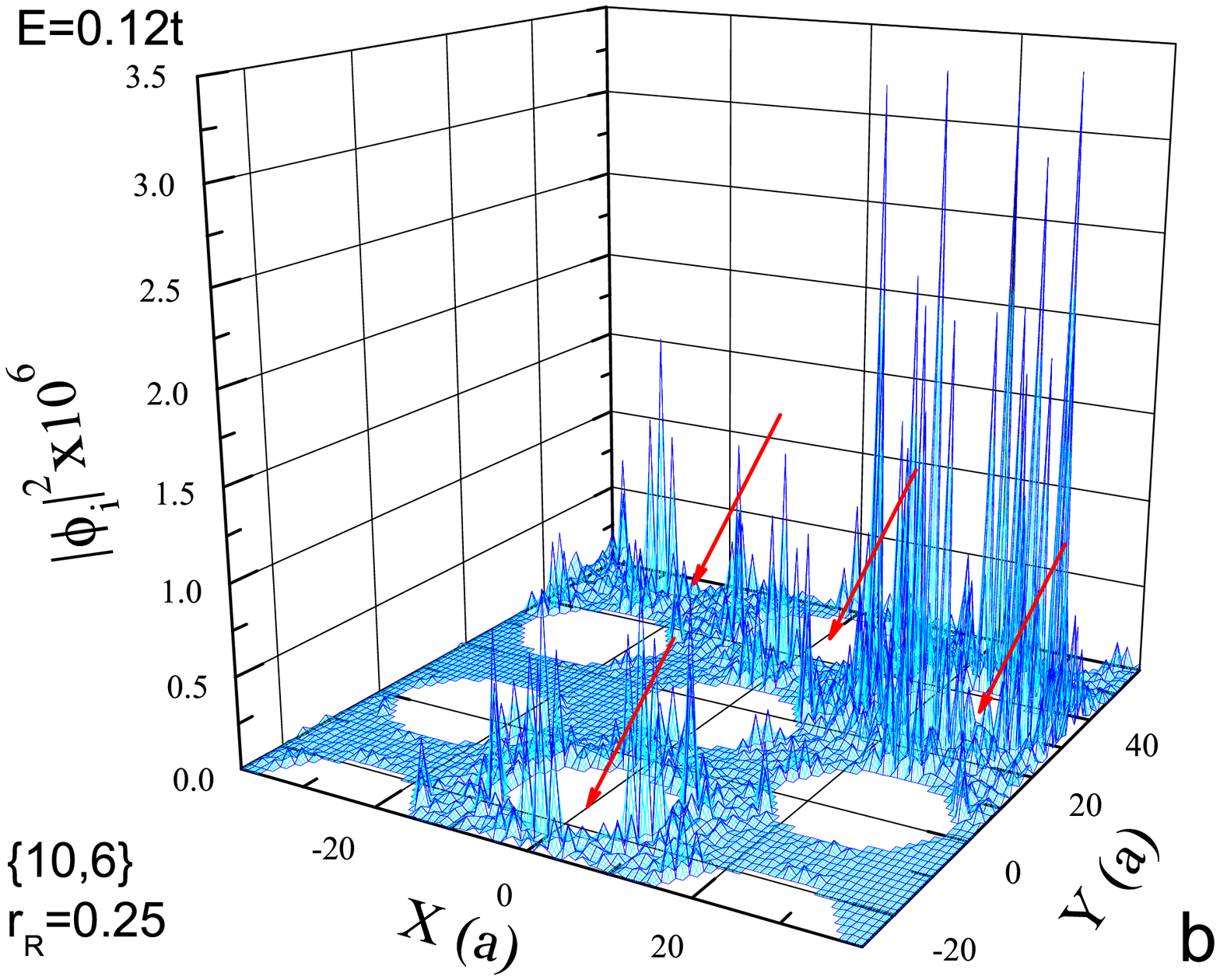}
}
\end{center}
\caption{Distribution of quasieigenstates of a $\{10,6\}$ GAL with geometrical disorder, where the radius of the holes is randomly shrunk or enlarged within the range $[5.75,6.25]$. The holes with unchanged radius ($R=6$) are indicated by the red arrows. The quasi-eigenstates are calculated at energy $E=0.056t$ and $E=0.12t$, corresponding to the states at the low energy peaks marked by the black and red arrows respectively in the DOS of Fig. \ref{Fig:GeomDisorder} b ($r_R=0.25$). The highest contribution to the quasi-eigensate at $E=0.056t$ (plot a) is concentrated around holes with radius $R>6$, for which the first ring of atoms, as compared to perfect GAL with $R=6$, has been removed. The quasi-eigenstate at $E=0.12t$ with highest amplitude (plot b) are localized around the holes with $R=6$ (marked by the red arrows), as in perfect GAL. }
\label{Fig:GeomQuasi}
\end{figure*}

In this section we present the results and discuss the effect of the different kinds of disorder introduced in Sec. \ref{Sec:Method}, in the DOS and in the optical conductivity of GALs. As discussed in Sec. \ref{Sec:Method} and sketched in Fig. \ref{Fig:Sketch}, we consider three main sources of disorder: {\it geometrical} disorder, which is associated to deviations of the GAL from the perfect periodicity; {\it resonant impurities}, which can be associated to additional vacancies in the graphene lattice, or to {\it adatoms} deposited on the sample; and the effect of on-site potentials which can randomly vary within the sample.

\subsection{Geometrical disorder}

We start by considering the most generic source of disorder in these kind of systems, which is the {\it geometrical} disorder. Uncontrollable fluctuations in the fabrication process  lead to irregularities in the resulting antidot lattice, such as changes in the center-to-center distance of the etched holes, or in variations in the size of the holes. Examples of the geometrical disorder in the lattice are sketched in Fig. \ref{Fig:Sketch} (a) and (b), respectively. Let us consider first the effect of a random deviation of the relative distance among the holes on the DOS and $\sigma(\omega)$, as shown in Fig. \ref{Fig:GeomDisorder}(a) and (c) respectively. In Fig. \ref{Fig:GeomDisorder}(a) we see that, for the perfect periodic array ($l_C=0$, black line) a clear band gaps open up in the spectrum. Notice that the peaked structure of the DOS is due to the set of locally flat bands which appear in the new band structure of the GAL as compared to the spectrum of standard graphene.\cite{PP08} If we now allow for a relative displacement among the nanoholes ($l_C\ne 0$), we observe that the gap shrinks but survives if $l_C$ is not too large, as seen by the red line of Fig. \ref{Fig:GeomDisorder} (a), but eventually the gap closes for some critical value of $l_C$, due to the lack of periodicity in the GAL, as it is the case shown by the green line of Fig. \ref{Fig:GeomDisorder} (a) which correspond to $l_C=a$. The disorder affects the optical conductivity in the following manner. As it can be seen in Fig. \ref{Fig:GeomDisorder} (c), for the perfect periodic case ($l_C=0$) $\sigma(\omega)=0$ up to $\omega=\Delta$, where $\Delta\approx 0.2t$ (for the case considered here)  is the gap opened due to the antidot array. Because $\Delta$ decreases as we increase $l_C$, the threshold for optical transitions is reduced and for $l_C=0.5a$ we observe a finite optical conductivity for $\omega\gtrsim 0.05t$. Finally $\sigma(\omega)>0$ at any frequency for an even larger amount of disorder as, e. g., $l_C=a$ (green line), for which the gap of the GAL has completely collapsed.

A similar effect on the electronic properties is observed if instead of randomly changing the relative separation between the antidots, their size is varied within some range, as sketched in Fig. \ref{Fig:Sketch} (b). The results of our simulations for this kind of disorder are shown in Fig. \ref{Fig:GeomDisorder} (b) and (d) for the DOS and optical conductivity, respectively. One observes that the DOS presents an increasing number of peaks as $r_R$ is increased. These peaks are associated to states with a large amplitude circling  the antidots, and their energy depends on the radius of the antidot. For the pristine GAL ($r_R=0$, black line) all antidots have the same radius which leads to the peaks at $E/t\sim \pm 0.12$ [signaled by a black arrow in Fig. \ref{Fig:GeomDisorder} (b)].  The finite width of the peak is due to a coupling between the antidots,  and the weak splitting reflects the van Hove singularities at the edges of these quasi-one dimensional bands. Peaks at higher energies originate from states that are not tightly localized around the antidots, but have a larger amplitude all over the sample. If the radius of the antidots is varied we observe that, apart from the peak discussed above shown by the black arrow, part of the spectral weight is transferred to new peaks that correspond to localized states at different energies, around antidots of different radii. Some examples are shown by the red and green arrows in Fig. \ref{Fig:GeomDisorder} (b). This behavior is illustrated by the spatial distribution of the quasi-eigenstates shown in Fig. \ref{Fig:GeomQuasi}. There we show, for the case of $r_R=0.25$, a small section of the lattice studied in our simulations, with a real space distribution of the amplitude of the quasi-eigenstates $|\Phi(E)|^2$. Fig. \ref{Fig:GeomQuasi} (a) shows that, for the energy $E/t\approx0.056$ [marked by the red arrow in Fig. \ref{Fig:GeomDisorder} (b)], the large amplitude of the states is  around antidots with $R>6$, for which the innermost ring carbon atoms of the antidot has been removed. However, in Fig. \ref{Fig:GeomQuasi} (b) we see that at an energy corresponding to the first mode of the undistorted lattice [$E/t\approx0.12$, shown by the black arrow in Fig. \ref{Fig:GeomDisorder} (b)] the states are localized at the edges of holes with a radius corresponding to perfect GAL. Those antidots are shown by red arrows in Fig. \ref{Fig:GeomQuasi}.

The effects of this kind of disorder on the optical conductivity are shown in Fig. \ref{Fig:GeomDisorder} (d). As $r_R$ is increased, we observe optical processes of lower and lower energy contributing to $\sigma(\omega)$, due to optical transitions between the localized states around holes of different sizes. This suggests that photoluminescence spectroscopy can be an useful tool for the characterization of the GALs.

\begin{figure*}[t]
\begin{center}
\mbox{
\includegraphics[width=0.8\columnwidth]{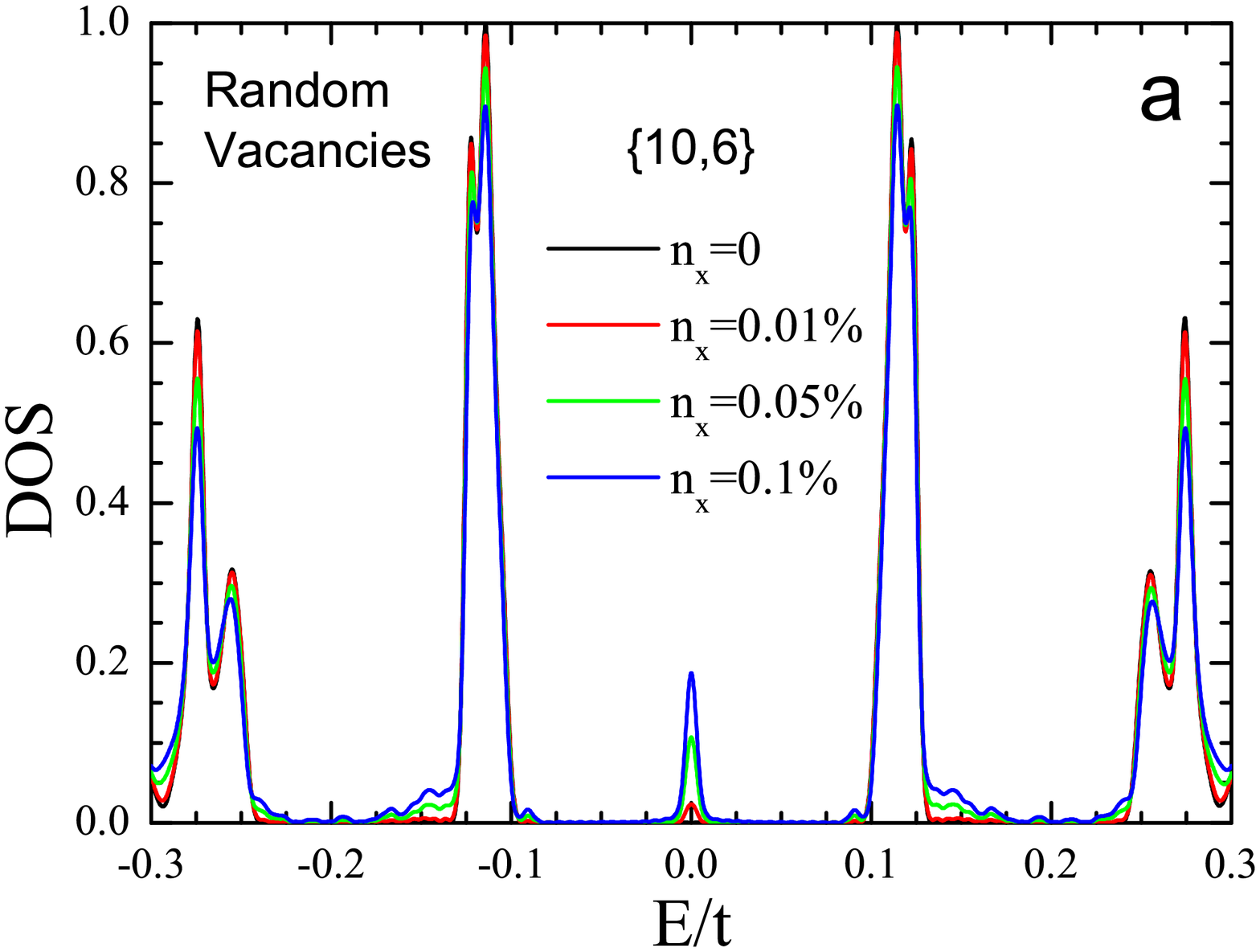}
\includegraphics[width=0.8\columnwidth]{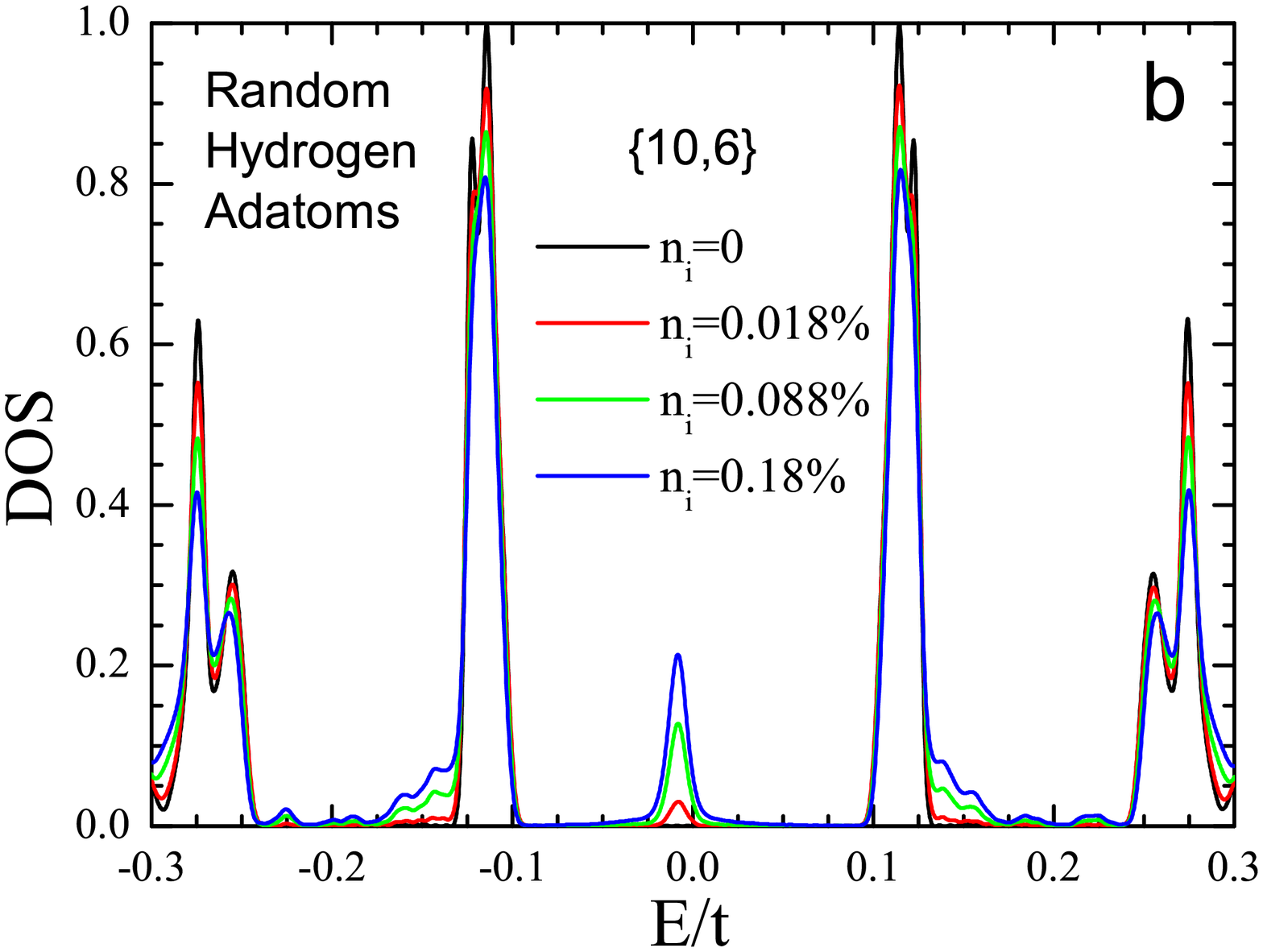}
}
\mbox{
\includegraphics[width=0.8\columnwidth]{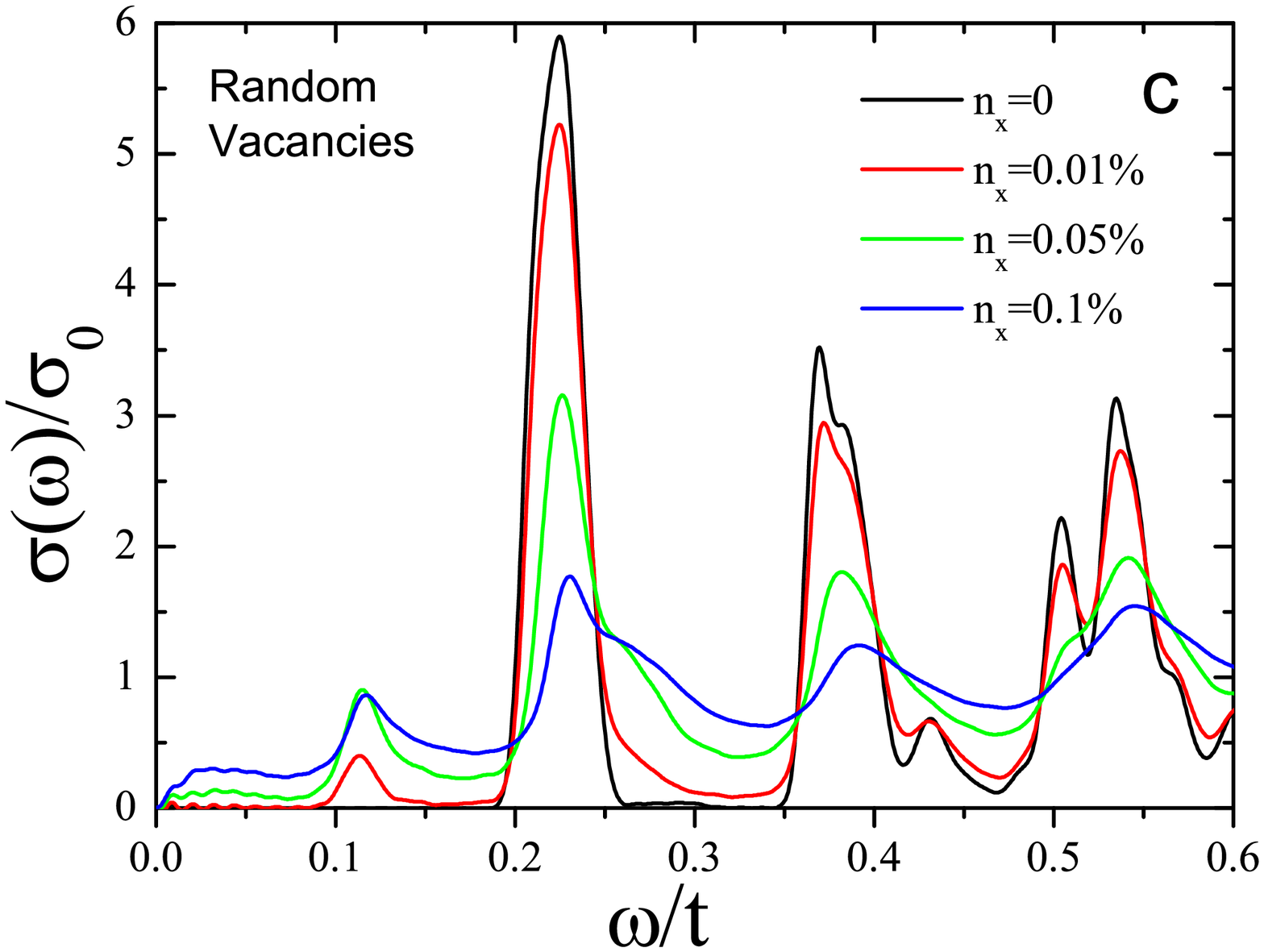}
\includegraphics[width=0.8\columnwidth]{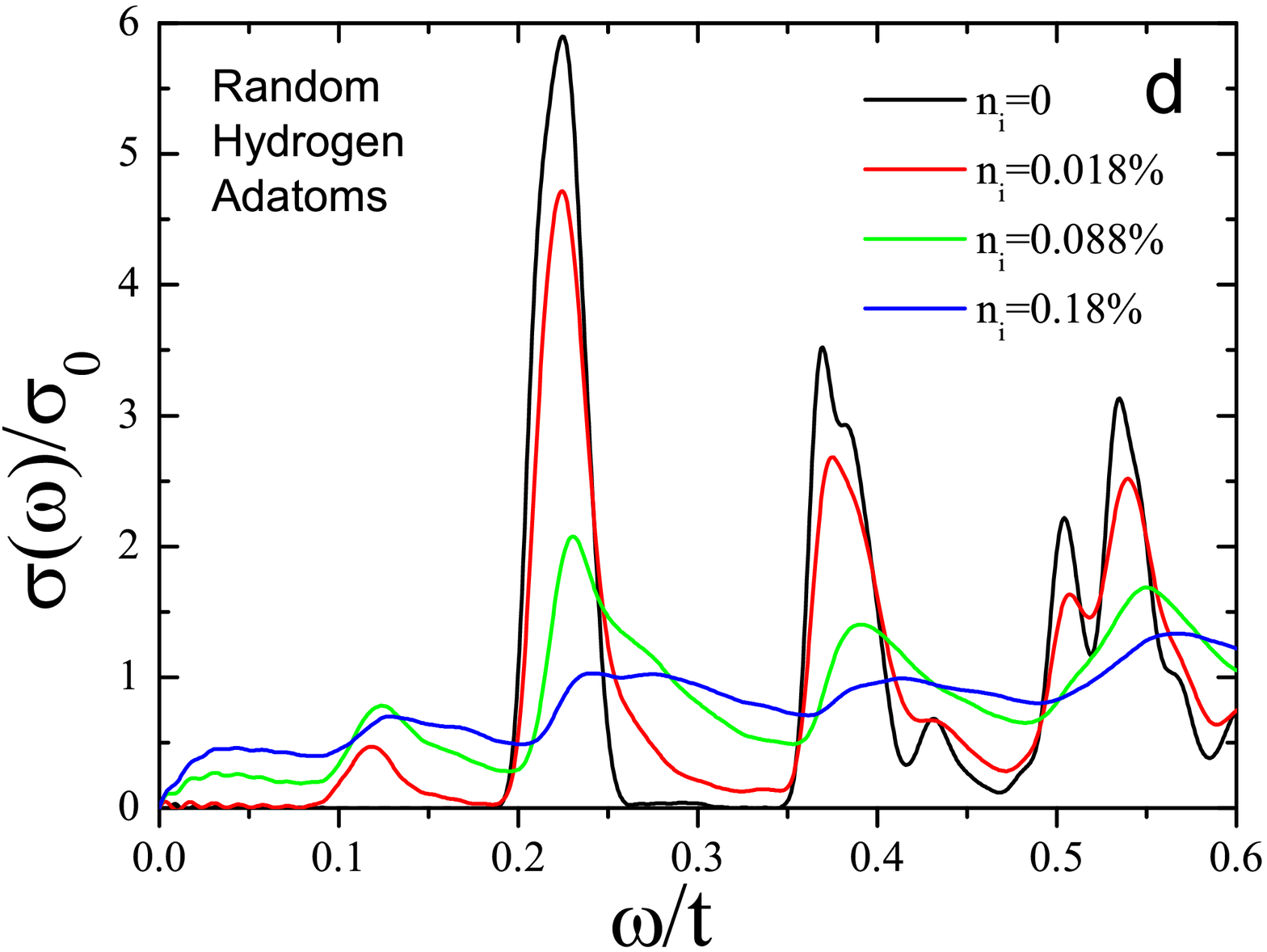}
}
\end{center}
\caption{DOS (top panels) and optical conductivity (bottom panels) for a $\{10,6\}$ GAL with resonant impurities. In the first column we show (a) DOS and (c) $\sigma(\omega)$ for a GAL with a random distribution of vacancies, as sketched in Fig. \ref{Fig:Sketch} (c). The different colors correspond to different amounts of missing dangling bonds, as denoted in the inset of the figures. In the second column we show (b) DOS and (d) $\sigma(\omega)$ for a GAL with hydrogen adatoms, as sketched in Fig. \ref{Fig:Sketch} (d). Different colors correspond to different percentage of adatoms in the sample.}
\label{Fig:ResonDisorder}
\end{figure*}

\subsection{Resonant impurities}

The next source of disorder that we consider is the effect of resonant scatterers. Resonant impurities can be understood as vacancy atoms in the sample, or as hydrogen or other organic molecules (CH$_3$, C$_2$H$_5$, etc) adsorbates which bind to a single carbon atom, changing its hybridization from $sp^2$ to $sp^3$ \cite {KatslesonBook,WK10}. A sketch of a GAL with a certain amount of vacancies or hydrogen adatoms randomly distributed is shown in Fig. \ref{Fig:Sketch} (c) and (d) respectively. The main effect of the resonant impurities in graphene membranes is the creation of "mid-gap" states at the Dirac point.\cite{WK10,YRRK11} Therefore, if some amount of these kind of impurities is present in the GAL, a zero energy flat impurity band is expected to appear in the middle of the gap. This is indeed what we obtain in our calculations, as it can be seen by the $E\approx 0$ peak in the DOS plots shown in Fig. \ref{Fig:ResonDisorder}. As we discussed in Sec. \ref{Sec:Method}, this kind of disorder is  accounted for in our calculations through the term ${\cal H}_{imp}$ in Eq. (\ref{Eq:Himp}), with the band parameters $V\approx 2t$ and $\varepsilon_d\approx -t/16$, as obtained from {\it ab initio} density-functional theory.\cite{WK10} The DOS of a GAL with different amounts of vacancies and hydrogen adatoms, randomly distributed, is shown in Fig. \ref{Fig:ResonDisorder} (a) and (b) respectively. We observe that, apart from a slight deviation from the Dirac point ($E=0$) of the position of the hydrogen adatoms impurity band (due to the finite value of the energy $\varepsilon_d$), as compared to the $E=0$ energy of the mid-gap band due to vacancies, the effect of these two kind of defects in the spectrum is very similar. In the two cases, the quasi-localization of the newly created states leads to an almost flat band which does not affect the rest of the energy spectrum away from the Dirac point (apart from some smearing of the peaks in the DOS).

As a consequence, the main contribution to the optical conductivity is obtained, as in the clean limit, for inter-band processes with an energy $\omega\approx\Delta$, as it can be seen in Fig. \ref{Fig:ResonDisorder} (c)-(d). However, due to the transfer of spectral weight to the mid-gap states, there is some finite $\sigma(\omega)$ for energies smaller than the threshold defined by the energy gap $\Delta$, with an appreciable peak at $\omega\approx \Delta/2\approx 0.1t$. This contribution is due to the new optical transitions from the impurity band to the conduction band, which are activated for $\omega>\Delta/2$.

\begin{figure*}[t]
\begin{center}
\mbox{
\includegraphics[width=0.8\columnwidth]{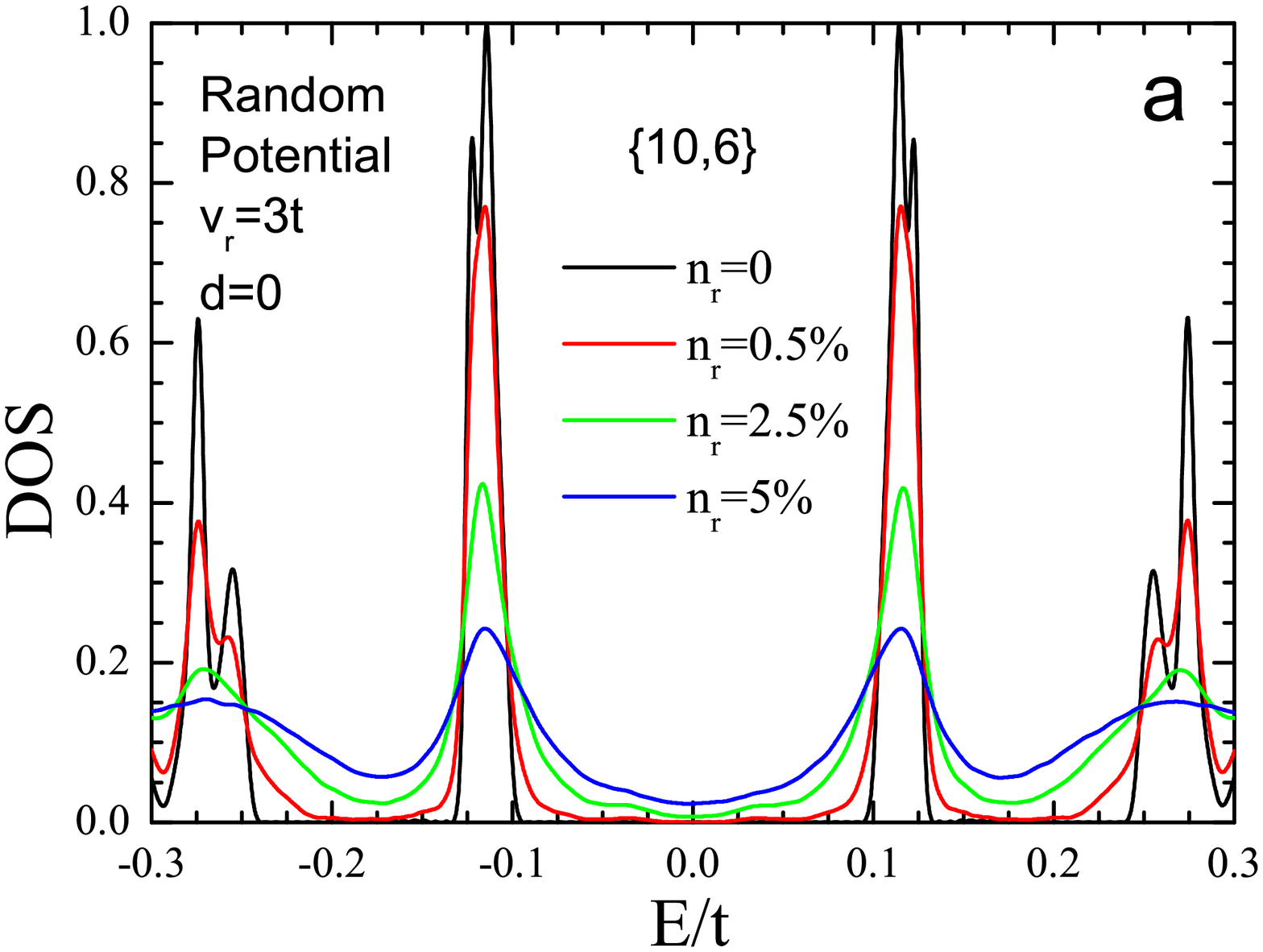}
\includegraphics[width=0.8\columnwidth]{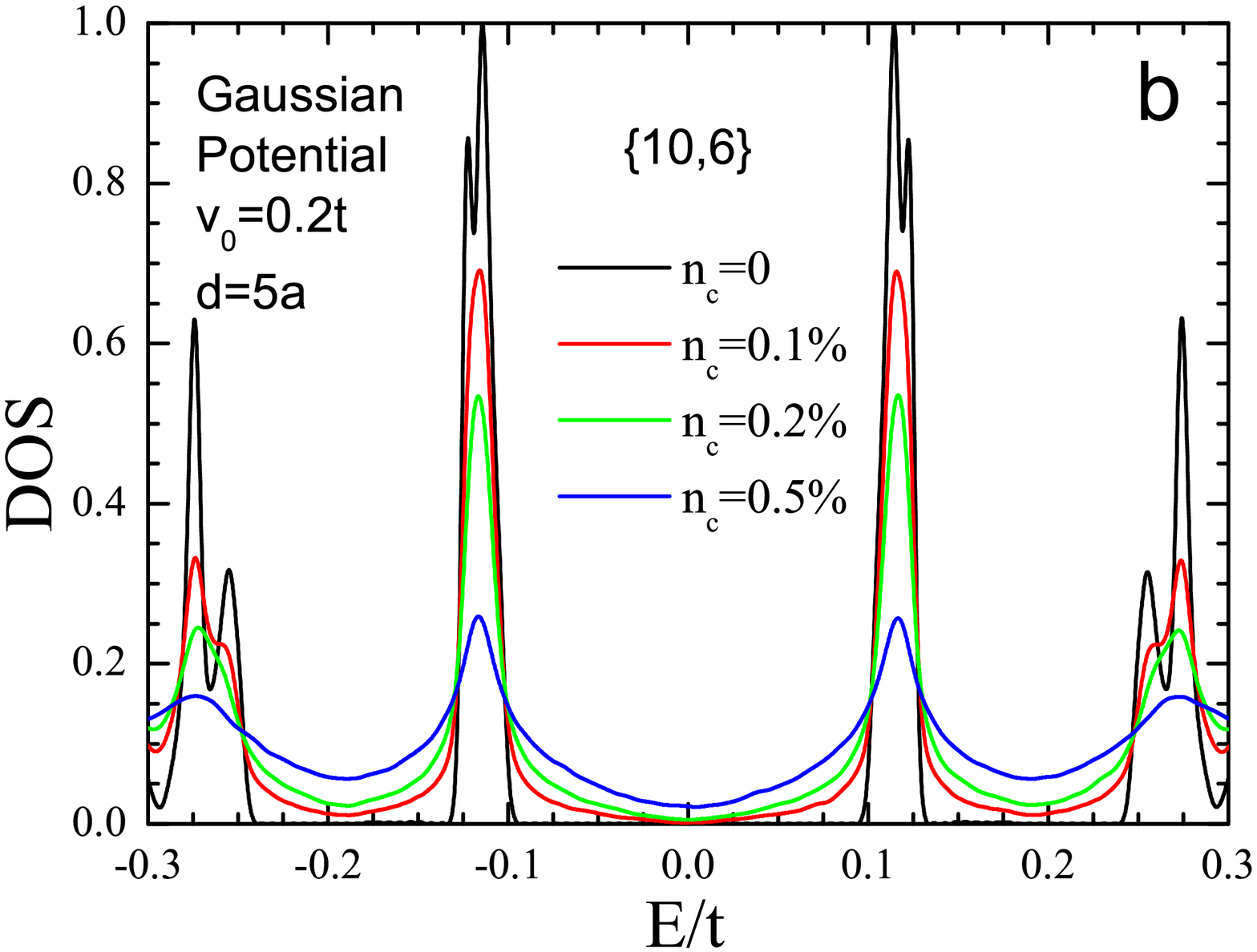}
}
\mbox{
\includegraphics[width=0.8\columnwidth]{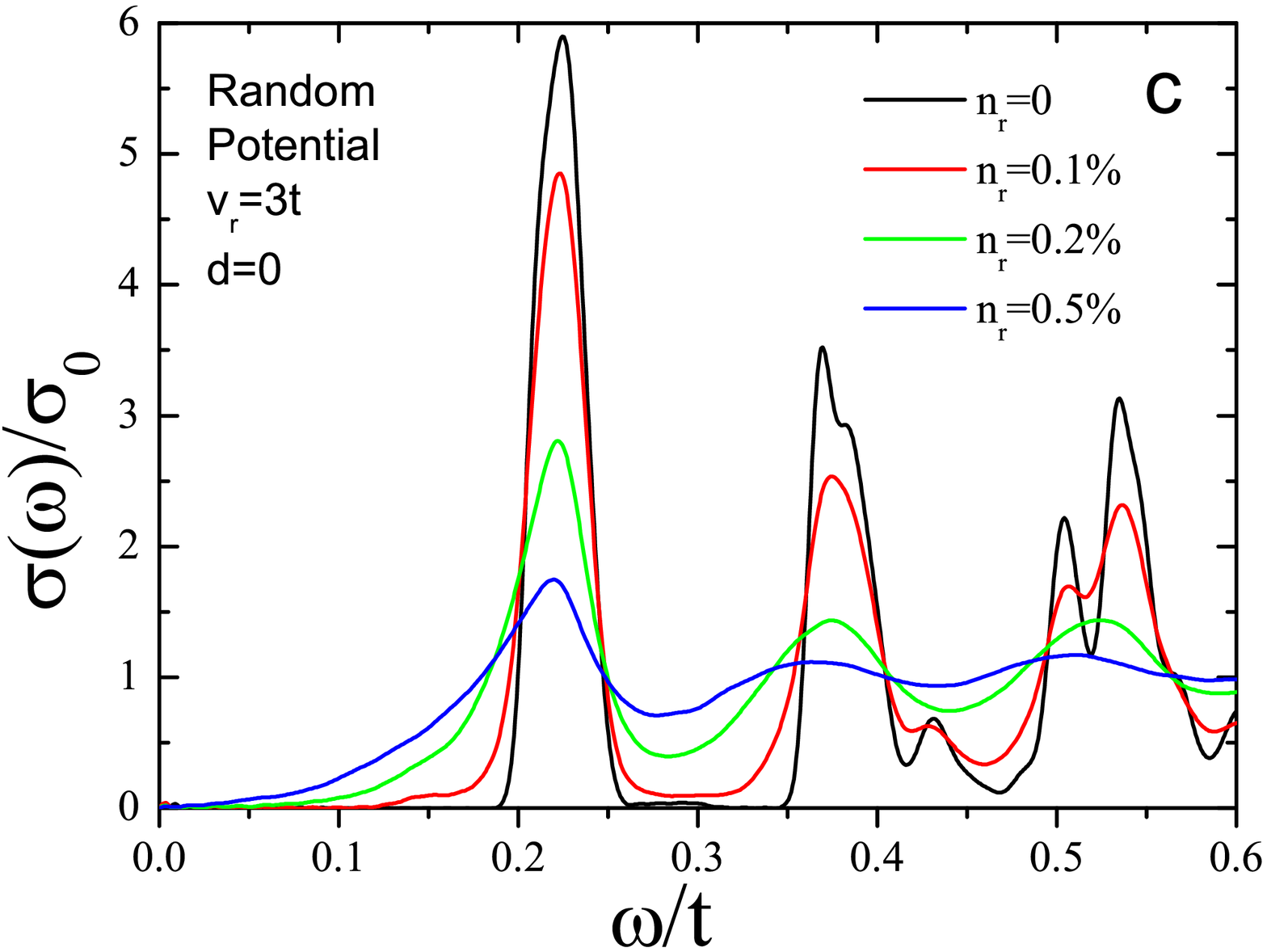}
\includegraphics[width=0.8\columnwidth]{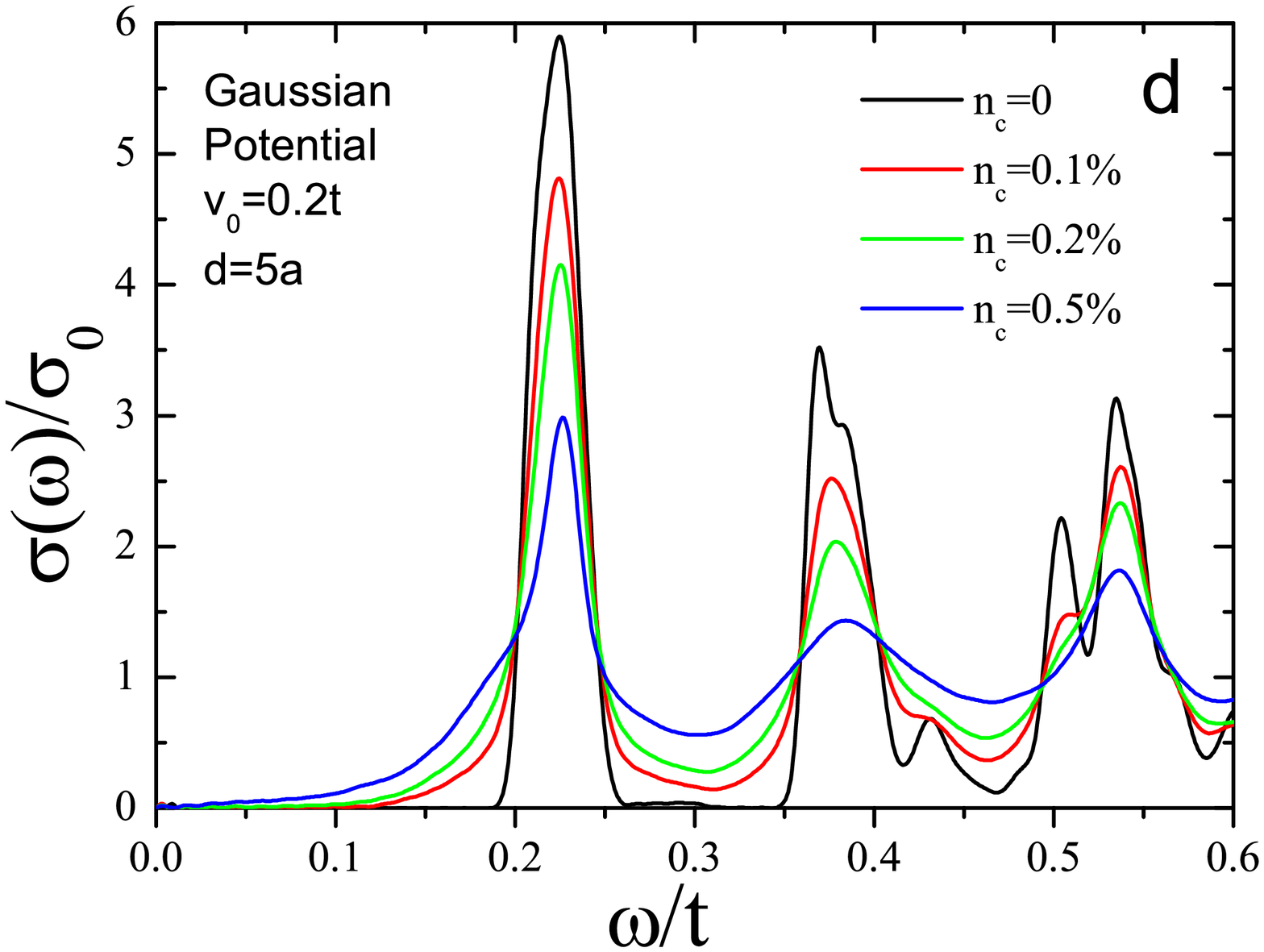}
}
\end{center}
\caption{DOS (top panels) and optical conductivity (bottom panels) for a $\{10,6\}$ GAL with on-site potential disorder. In the first column we show (a) DOS and (c) $\sigma(\omega)$ for a GAL with a non-correlated random distribution of short-range potential, which can take the values within the range $[-v_r,v_r]$. The different colors correspond to different concentrations of disorder, as denoted in the inset of the figures. In the second column we show (b) DOS and (d) $\sigma(\omega)$ for a GAL with a long-range Gaussian potential disorder. The potential is given by Eq. (\ref{Eq:Gaussian}), where $V_k$ is uniformly random in the range between $-V_0$ and $V_0$, and $d$ is the effective potential radius. (See text).
}
\label{Fig:PotDisorder}
\end{figure*}

\subsection{Short- and long-range potential disorder}

Another kind of disorder that can be considered is a shift of the on-site potentials at a given lattice points, which can lead to a local shift of the chemical potential. This contribution is accounted for by means of the second term of the Hamiltonian (\ref{Eq:H0}). This kind of disorder can be of extraordinary importance. For example, if the atoms in sublattices A and B have opposite strength of the on-site potential $v_r$, then a gap of size $\Delta=2v_r$ is opened in the spectrum.\cite{YRK10} Here we consider, depending on how the defects are distributed over the lattice sites, a correlated or a non-correlated disorder. In the case of a short-range and non-correlated potential disorder, the nonzero on-site potentials are taken to be uniformly randomly distributed over the sample within a range $[-v_r,v_r]$. The results for the DOS and the optical conductivity with $v_r=3t$ are shown in Fig. \ref{Fig:PotDisorder} (a) and (c), respectively. We observe a broadening of the peaks in the DOS, accompanied by a transfer of spectral weight to the gaped regions.

Next, consider the long-range correlated disorder given by Eq. (\ref{Eq:Gaussian}).
In standard graphene, this kind of disorder leads to regions of the graphene membrane where the Dirac point is locally shifted to the electron ($V_k <0$) or to the hole ($V_k > 0$) side with the same probability. This leads to some finite DOS at zero energy. Our calculations for GALs in the presence of a Gaussian potential disorder are shown in Fig. \ref{Fig:PotDisorder} (b) and (d), for the DOS and optical conductivity respectively. One observes similar qualitative effects in the spectra as compared to the short-range non-correlated random potentials. In particular, there is a small but appreciable contribution to the optical conductivity at low frequencies, due to the transfer of states to the gapped region.

\section{Conclusions}\label{Sec:Conclusions}

In conclusion, we have presented a systematic study of the effect of disorder in GALs. We have used a tight-binding model in a perforated honeycomb lattice of carbon atoms. The DOS has been calculated from a numerical solution of the TDSE, whereas the optical conductivity has been obtained by using the Kubo formula for non-interacting electrons. We have considered the most generic sources of disorder in these kind of samples: geometrical disorder such as random deviation of the periodicity and of the radii of the nanoholes from the perfect array, as well as the effect of resonant scatterers in the sample (e.g., vacancies, adatoms, etc.) and the effect of non-correlated and correlated (Gaussian) on-site potentials.
In order to have a qualitative understanding of the effect of the different kinds of disorder on the samples, we have applied the method to one representative case, namely a $\{10,6\}$ GAL. However, we emphasize that this scheme is completely general and applicable to any set of parameters $\{L,R\}$.

Our results show that the gap is rather robust against geometrical disorder, and only a large deviation of the antidot array from the perfect periodicity leads to a narrowing and eventually closing of the energy gap. We obtain localized states encircling the antidots, the energy of which depends on the radius of the hole. The presence of additional resonant scatterers, as vacancies or adatoms, leads to the creation of midgap states. The existence of this impurity band is reflected in the optical conductivity, which now extends to energies smaller than the gap energy $\Delta$, due to disorder activated optical transitions from the impurity band to the conduction band. However, the main contribution to $\sigma(\omega)$ still corresponds to transitions with an energy of the order of $\Delta$. Finally, the presence of non-correlated or of Gaussian potential disorder leads to a smearing of the peaks in the DOS, as well as to the transfer of spectral weight to the gapped region. Contrary to the effect of resonant scatterers, the presence of potential disorder does not create a zero-energy band with a prominent peak in the DOS at $E=0$, but instead a DOS that grows smoothly as a function of energy within the gapped region. As a consequence, the optical conductivity also grows slowly from 0 until it reaches its maximum contribution at the energy of the gap. Therefore, photoluminescense spectroscopy experiments could be useful for the characterization of the GALs.

\section{Acknowledgments}

We thank Thomas Garm Pedersen for valuable discussions. The support by the Stichting Fundamenteel Onderzoek der Materie (FOM) and
the Netherlands National Computing Facilities foundation (NCF) are
acknowledged. We thank the EU-India FP-7 collaboration under MONAMI. RR acknowledges financial support from the Juan de la Cierva Program (MEC, Spain).
The Center for Nanostructured Graphene CNG is sponsored by the Danish National Research Foundation.

\bibliographystyle{apsrev}
\bibliography{BibliogrGrafeno_APJ}

\end{document}